\documentclass[amsfonts,nofootinbib,prd,aps]{revtex4}

\usepackage{graphicx}

\begin{document}

\title{Self-adjointness in the Hamiltonians of deparameterized totally
  constrained theories: a model}

\author{ Rodolfo Gambini$^{1}$, Jorge Pullin$^{2}$} 
\affiliation { 1. Instituto de F\'{\i}sica,
  Facultad de Ciencias,
  Igu\'a 4225, esq. Mataojo, Montevideo, Uruguay. \\
  2. Department of Physics and Astronomy, Louisiana State University,
  Baton Rouge, LA 70803-4001}

\begin{abstract}
  Several proposals to deal with the dynamics of general relativity
  involve gauge fixings or the introduction matter fields in terms of
  which the theory is deparameterized. The resulting theories have
  true Hamiltonians for their evolution that usually involve square
  roots, and this poses certain challenges for their implementation as
  self-adjoint quantum operators. We show in the context of a simple
  model of totally constrained theory that one can introduce related,
  well defined operators that reproduce semiclassically the same
  physics as the original ones, at least for states peaked in the
  regions of phase space where their associated classical quantities
  are well defined.
\end{abstract}

\maketitle
\section{Introduction}

Two important problems in the canonical approach to quantum gravity
are the issue of the constraint algebra and of the physical
interpretation of the dynamics of the theory. In the usual canonical
formulations, general relativity has constraints that close a first
class algebra in the Dirac sense. It is not a Lie algebra as it has
structure functions. In particular the Poisson bracket of two
Hamiltonian constraints is proportional to a diffeomorphism constraint
with the proportionality factor a function of the canonical
variables. This poses problems at the time of quantizing the
theory. In particular it is easy to show that one cannot implement the
constraints as self adjoint operators and hope to close the algebra at
a quantum level. The physical interpretation of the dynamics of the
theory also requires to disentangle the resulting ``frozen''
formalism, since there is no non-vanishing true Hamiltonian for the
theory, in terms of Dirac observables. However, for the case of pure
gravity in vacuum we do not know a single Dirac observable in closed
form (apart from certain formal constructions \cite{stornaiolo}). One
can expect that approximate treatments may yield expressions for Dirac
observables, but only in certain regimes \cite{dittrich}.

The previously listed problems have led to various proposals to deal
with them. One possibility is to gauge fix the theory.  Gauge fixing
eliminates the constraints, the issue of the constraint algebra, and
reveals the true dynamics of the theory. The quantization should
therefore presumably be straightforward. But it turns out it is
not. An example is given by recent proposals to gauge fix spherically
symmetric gravity coupled to a scalar field \cite{nestor}. Although a
gauge fixing in which the true Hamiltonian is the integral of a local
scalar density has recently been achieved, its definition involves
square roots. In this particular case there is no reason why
generically what appears under the square root should be a positive
quantity. This just shows that one cannot find a single gauge fixing
that works generically for all types of initial data (in the proposal
of \cite{nestor} one can find different gauge fixings for different
types of initial data that yield positive quantities inside the square
roots). This example is particularly revealing of the limitations one
faces when gauge fixing since in it all things are analytically  under
control. Nevertheless, promoting the resulting Hamiltonian to a
self-adjoint operator is challenging due to the presence of square
roots.

Another proposal is to consider the inclusion of matter in the theory
as a tool to deparameterize it and construct Dirac observables,
starting with the pioneering work of Brown and Kucha\v{r} \cite{brku}
(see \cite{thgi} for a recent review). In a sense, this is very
natural, since in the real world we rely on matter to construct
reference frames in terms of which we describe physics. With the
introduction of a certain kind of matter, one can use it as as clock to
deparameterize the theory and have a true Hamiltonian. In many of
these proposals the diffeomorphism constraint remains, but the true
Hamiltonian is diffeomorphism invariant. However, since most forms of
matter (see \cite{thiemannk} for an exception) have Hamiltonians that
are quadratic in the fields and momenta, when one uses them to
deparameterize the theory one is again led to expressions for the
Hamiltonian that involve square roots. One can show in many cases that
the expression under the square root is classically positive definite
on-shell. However, when one quantizes there have been arguments put
forward \cite{asho,haji} saying that one needs to consider negative values.

Motivated by these occurrences of square roots, we would like to
analyze a simple model where square roots occur, and to study the
consequences of one procedure to deal with them upon
quantization. The model is the one considered by Rovelli
\cite{rovelli} and can be thought of as two harmonic oscillators with
a constant sum of their energies. The quantization of this model is
challenging since the phase space is compact and it therefore does not
admit a Hamiltonian structure. Nevertheless global Dirac observables
exist and a full quantization is possible \cite{GaPo}. It is therefore
a good arena to compare the full quantization with the types of
``truncated'' quantizations obtained by using the techniques one uses
to deal with square roots.

\section{The model and its  quantization}
The model \cite{rovelli} has a four dimensional phase space $q_1, q_2,
p_1, p_2$ with a constraint given by
\begin{equation}
C= \left[-\frac{1}{2}\left(p_1^2+p_2^2+q_1^2+q_2^2\right)+M\right],
\end{equation}
with $M$ a constant. The surface $C=0$ is a pre-symplectic space
without a true Hamiltonian structure since it is compact. 

The reduced phase space due to the constraint $C=0$ can be
parameterized as,
\begin{eqnarray}
  q_1 &=& \sqrt{2A}\sin \tau,\label{2}\\
  q_2 &=& \sqrt{2M-2A}\sin\left(\tau + \phi\right).\label{3}
\end{eqnarray}
$A$ and $\phi$ constants that can later be identified as Dirac
observables of the theory.  If we choose the lapse  as $N=1$ then
$\dot{q}_1=-p_1$ and $p_1=-\sqrt{2A}\cos\tau$ and similarly
$\dot{q}_2=-p_2$ and $p_2=-\sqrt{2M-2A} \cos\left(\tau + \phi\right)$.

The exact theory for this model was developed in \cite{GaPo}. One can
define\footnote{The definition chosen differs from that of
  \cite{rovelli} and \cite{GaPo}. What we call $L_y$ would correspond
  to $L_x$ of those references and what we call $L_x$ would be
  $-L_y$. Both satisfy the algebra of angular momenta. The truncated
  theory we will consider later does not have such symmetry and only
  approximates well the choice we make in this paper.}  three angular
momentum Dirac observables,
\begin{eqnarray}
  L_x &=& -\frac{1}{2}\left(p_1q_2-p_2q_1\right)=
-\sqrt{A \left(M-A\right)}\sin \phi,\label{4}\\
  L_y &=& \frac{1}{2}\left(p_1p_2+q_1q_2\right)=
\sqrt{A \left(M-A\right)}\cos \phi,\\
  L_z &=& \frac{1}{4}\left(p_1^2-p_2^2+q_1^2-q_2^2\right)=A-\frac{M}{2},
\end{eqnarray}
such that $L^2 = \frac{M^2}{4}$. One can define an angular momentum
basis $\vert j, m\rangle$ with $j$ integer or half-integer such that
$\hat{L}_z\vert j, m\rangle = m \vert j, m\rangle$. Notice that in the
constant $M$ cannot take arbitrary values in the quantum theory,
$M^2_{\rm Full}=j(j+1)$. We use ``full'' to refer to the full
quantization since later we will compare with the truncated
quantization.

The exact expression for the observable $\hat{A}$ is given by,
\begin{equation}
  \hat{A}_{\rm Full}\vert j, m\rangle = \left(\hat{L}_z +
    \frac{M}{2}\right)\vert j, m \rangle = \left( m
    +\sqrt{j(j+1)}\right)\vert j, m\rangle. 
\end{equation}

For the observable $L_x$, $L_x$ we introduce the usual notation of
raising and lowering operators $L_x=(L_++L_-)/2$ and
$L_y=(L_+-L_-)/(2i)$, in terms of which we have,
\begin{equation}
  \hat{L}^{\rm Full}_x\vert m\rangle=\frac{\hat{L}_++\hat{L}_-}{2} \vert
  m\rangle=
\frac{\sqrt{j(j+1)-m(m+1)}}{2} \vert j,m+1\rangle+
\frac{\sqrt{j(j+1)-m(m-1)}}{2} \vert j,m-1\rangle,
\end{equation}
and we will find convenient to compare with the truncated theory to 
redefine $m=j-n$, for a given $j$,
\begin{equation}
\hat{L}_x^{\rm Full}\vert j,m\rangle =
\hat{L}_x^{\rm Full}\vert n\rangle =
\frac{\sqrt{(n+1)(2j-n)}}{2}\vert n+1\rangle
+
\frac{\sqrt{n(2j-n+1)}}{2}\vert n-1\rangle.
\label{9}
\end{equation}


A similar construction can be carried out for $\hat{L}_y^{\rm Full}$.
\section{Gauge fixing}

Let us now consider a gauge fixed treatment of the model. To this aim
we will proceed locally in phase space, and introduce a total
Hamiltonian $H_T=N C$ with $N$ a Lagrange multiplier in order to fix
the gauge. We choose a gauge $q_1=t$, with $t$ the time
parameter associated with the evolution generated by the total
Hamiltonian. This leads to $\dot{q}_1=1$ and therefore,
\begin{equation}
  1=\left\{q_1, H_T\right\}=-N p_1,
\end{equation}
which fixes the lapse $N=-1/p_1$ with $p_1=-\sqrt{2 M -p_2^2
  -q_2^2-t^2}$ and therefore $N>0$. This already tells us that the
Hamiltonian theory is not globally defined as for large enough $t$ the
square root is imaginary. The total Hamiltonian now reads, 
\begin{equation}
  H_T=\frac{1}{ p_1}
  \left[\frac{1}{2}\left(p_1^2+p_2^2+q_1^2+q_2^2\right), \label{11}
-M\right]
\end{equation}
before the strong imposition of the constraint. We get the equations
of motion by computing the Poisson brackets of the variables with the
total Hamiltonian,
\begin{eqnarray}
  \dot{q}_2&=&\frac{p_2}{p_1},\\
  \dot{p}_2&=&-\frac{q_2}{p_1}.
\end{eqnarray}
These equations can be obtained from a true Hamiltonian,
\begin{equation}
  H_{\rm True}=\sqrt{2 M -p_2^2-q_2^2-t^2}.\label{14}
\end{equation}

Due to the square root, at a quantum level $H_{\rm True}$ will not
become a self-adjoint operator. However, one can define Hamiltonians
that approximate well the exact solutions for semiclassical
excitations around classical exact solutions for which $H_{\rm True}$
is real. An example could be to take the absolute value of what is
inside the square root in $H_{\rm True}$. Or to consider its square
and then take the real branch of its fourth root. If one thinks of
cases of interest, like gauge fixings in spherically symmetric
gravity, this means one could use these techniques to study, for
instance, black hole evaporation for large black holes.

The exact evolution (\ref{2},\ref{3}) written in this gauge, 
since $q_1=t$, and $\tau=\sin^{-1}\left(\frac{t}{\sqrt{2 A}}\right)$, 
is given by 
\begin{eqnarray}
  q_2(t)&=&
\sqrt{\frac{M}{A}-1}\left[t\cos\phi
  +\sqrt{2A-t^2}\sin\phi\right]\\
p_2(t)&=& 
\sqrt{\frac{M}{A}-1}\left[\sqrt{2A-t^2}\cos\phi -t \sin\phi\right].
\end{eqnarray}
This solution can also be obtained by integrating the evolution
equation stemming from the Hamiltonians $H_T$ (\ref{11}) and $H_{\rm
  True}$ (\ref{14}).  Notice that the gauge fixed solution is not
globally defined as is readily seen from the presence of the inverse
trigonometric function, so not all values of $\tau$ are obtained from
$t$. The solution therefore covers a portion of phase space until for
some values of $t$ the solution becomes complex.

\section{Quantization of the truncated theory}

The quantization of the Hamiltonian $H_{\rm True}$ (\ref{14}) 
has the problem of the square root. In fact, one can
show \cite{haji} that the operators obtained by a straightforward
quantization of $H_{\rm True}$ are not normal. The strategy will
to substitute another expression for $H_{\rm True}$ such that they
both coincide in the region of the phase space where the argument of
the square root is positive, for instance \cite{Thiemann},
\begin{equation}
\tilde{H}_{\rm True} =\sqrt{\vert 2 M -p_2^2 -q_2^2 -t^2\vert} 
=\left[\left(2 M -p_2^2 -q_2^2 -t^2\right)^2\right]^{1/4}.
\end{equation}
This Hamiltonian ensures that the equations of motion reproduce those
of the classical theory in the region in which $A>t^2/2$. So
$\tilde{H}_{\rm True}$ and $H_{\rm True}$ lead to the same solutions
in the region in which $H_{\rm True}$ is real. We call the resulting
theory ``truncated'' since it will differ from the original one for
large values of $p_2^2+q_2^2$. 

In order to quantize we notice that $p_2^2+q_2^2$ is the Hamiltonian of a
harmonic oscillator, and therefore one can use the quantization
technique of creation and annihilation operators. In particular the
Hamiltonian will be a function of the number operator. 

As usual we define the classical quantities,
\begin{eqnarray}
  a &=&\frac{1}{\sqrt{2}}\left(q_2+i p_2\right)\\
  a^*&=& \frac{1}{\sqrt{2}}\left(q_2-i p_2\right).
\end{eqnarray}
These quantities can be readily quantized. We introduce the number
operator $\hat{N}=\hat{a}^\dagger\hat{a}$, so we have,
\begin{equation}
  \hat{p}_2^2+\hat{q}_2^2=2 \hat{a}^\dagger\hat{a}+1 =2 \hat{N}+1,
\end{equation}
and we have that $\left[\hat{N},\hat{H}_{\rm
    True}\right]=0$. Introducing the number basis $\hat{N}\vert
n\rangle = n \vert n \rangle$, we have that
\begin{equation}
2 \hat{A}^{\rm Truncated}=  2 M_{\rm Truncated} -\hat{p}_2^2-\hat{q}_2^2 = 2 M_{\rm Truncated} -2 \hat{N}-1,
\end{equation}
with the Dirac observable becoming the self-adjoint operator,
\begin{equation}
  \hat{A}^{\rm Truncated}\vert n \rangle= 
\left(M_{\rm Truncated}-n -\frac{1}{2}\right)\vert n\rangle
  = A_n \vert n \rangle.
\end{equation}
Notice that in this quantization the value of $M_{\rm Truncated}$ is arbitrary, unlike
in the full quantization.

\section{Comparison of the full and truncated theories}

The Hilbert space of the truncated theory $\vert n\rangle$ with $n\in
[0,\infty]$ is infinite dimensional. The Hilbert space of the full
theory $\vert j,m\rangle$ with $j$ either a given integer or
semi-integer and $-j<m<-j$ and $m$ differing from $j$ by an integer is
finite dimensional with dimension $2j+1$. However, it is clear that if
one admits arbitrary values of $n$ in the truncated theory this will
not correspond to real solutions of the theory one started from since
$A<0$ in that case. To compare the full and truncated theories we need
to identify a correspondence between their Hilbert spaces. The best
way to see the correspondence is to identify $\vert j, m\rangle$ with
$\vert n\rangle$ with $n=j-m$. We will see also that $M_{\rm
  Truncated} =2j +1$ in order to reproduce the eigenvalues of
$\hat{A}_{\rm Full}$.

We would like to compare the Hilbert space of $\hat{\tilde{H}}$ with
that of the full theory. We will see  that ${\cal H}_{\rm Full} \subset
{\cal H}_{\rm Truncated}$. Since all quantities of the theory can be
written in terms of the Dirac observables, it suffices to study their
action.  We will see that their action coincides for $t^2/2<
A_n$.

Therefore in the truncated theory $M/2=j+1/2$.
Let us start with the observable $A$ 

The exact expression for the observable $\hat{A}$ in the full theory
is given by,
\begin{equation}
  \hat{A}_{\rm Full}\vert j, m\rangle = \left(\hat{L}_z +
    \frac{M}{2}\right)\vert j, m \rangle = \left( m
    +\sqrt{j(j+1)}\right)\vert j, m\rangle =A^{\rm Full}_{j,m} \vert j, m\rangle, 
\end{equation}
whereas the truncated expression is given by 
\begin{equation}
  \hat{A}_{\rm Truncated}\vert n \rangle = \left(M_{\rm Truncated} -n
    -\frac{1}{2}\right)\vert n \rangle =A^{\rm Truncated}_n \vert n \rangle. 
\end{equation}

The difference in eigenvalues of $\hat{A}_{\rm Exact}$ and
$\hat{A}_{\rm Truncated}$, using the identification of the
Hilbert spaces and the choice of $M_{\rm Truncated}= 2j+1$ is,
\begin{equation}
A^{\rm Full}_{j,j-n}-A^{\rm Truncated}_n=
  j +\sqrt{j(j+1)}-M +\frac{1}{2}=j +\sqrt{j(j+1)}-2 j -\frac{1}{2}.
\end{equation}
If $j\gg 1$ then the difference vanishes, it goes as $O(1/j)$. The
condition $A>0$ is equivalent to $2j-n>0$, so $n<2 j$ and the
corresponding subspace of the Hilbert
space has the same number of elements as in the full case.

To compare the second observable, let us consider $L_x$ in the
Heisenberg representation. In the gauge considered we have that
$q_1=t$, and $q_2, p_1, p_2$ are given in section III. Substituting
them in $L_x$ one recovers expression (\ref{4}),
\begin{equation}
  L_x =-\frac{1}{2} \left(p_1 q_2 - p_2 q_1\right)=
-\sqrt{A \left(M-A\right)}\sin \phi
\end{equation}
and is time independent and therefore its operator representations in
the Heisenberg and  Schr\"odinger representations coincide. This
expression can be rewritten as,
\begin{equation}
  \hat{L}_x=-\frac{1}{2}\sqrt{2\hat{A}^{\rm Truncated}}\hat{q}_2(0). 
\end{equation}
This can be realized in the basis $\vert n\rangle$ by substituting
$\hat{q}_2(0)$ in terms of the creation and annihilation operators. To
have a self-adjoint operator we write,
\begin{equation}
  \hat{L}_x=\frac{1}{2}\sqrt[4]{\hat{A}^{\rm Truncated}}
\left(\hat{a}+\hat{a}^\dagger\right) \sqrt[4]{\hat{A^{\rm Truncated}}},
\end{equation}
which explicitly gives,
\begin{equation}
    \hat{L}_x^H \vert n \rangle = 
\frac{\sqrt{(n+1)(2j-n)-\frac{1}{4}}}{2}\vert n+1\rangle+
\frac{\sqrt{n(2j-n+1)-\frac{1}{4}}}{2}\vert n-1\rangle. \label{29}
\end{equation}
We can now compare the action of this truncated operator with that of
the full theory, which we computed in equation (\ref{9}). The
semiclassical approximation works best when $A\gg t^2/2$ so we are away
from the place where the gauge fixing fails and when $A\ll M$ since the
expressions obtained were up to order $1/M$. In that regime equations
(\ref{9}) and (\ref{29}) agree. A similar discussion holds for
$\hat{L}_y$.

One can also compute the evolving constants $q_2(t),p_2(t)$ in this space
and compare with the exact ones of reference \cite{GaPo}. In the
regime discussed they agree.

\section{Conclusions}

We have shown in a totally constrained model that gauge fixing leads
to expressions that may be ill defined for certain regions of phase
space. This is in analogy with what occurs in gauge fixings in gravity
and when one introduces matter to deparameterize the theory. We show
that one can introduce a quantization based on a ``truncated'' version
of the theory with well defined self-adjoint operators. It reproduces
in the semi-classical limit the correct physics of the original theory
in the region of phase-space where the gauge fixing is well defined.

Among the lessons learned from the model is that the Hilbert space of
the truncated and full theory can be quite different and one needs to
restrict the one of the truncated theory in order to have agreement
between them. We also see one does not recover everything of the full
theory. In this particular case one of the constants of the model
takes a restricted set of values in the full theory and this is not
captured by the truncated theory. The approximation gets worse as one
gets close in phase space to where the gauge fixing stops being
valid. 

Another point is that we have considered the observables of the
truncated theory for positive $A$. For negative values of $A$ there
exist observables of the truncated theory that coincide for positive
values of $A$ with those of the full theory. For negative values of
$A$ the comparison makes no sense as the observables of the full
theory are not well defined. In particular the algebra of angular
momentum does not hold for those observables of the truncated
theory. That is the root of having to choose one particular form of
the observables of the full theory to approximate, since one does not
have present the symmetry $L_x\to L_y$, $L_y\to-L_x$ one has in the
angular momentum algebra.

This example suggests a procedure to extract (at least certain) physical
predictions from theories that cannot themselves be quantized properly
due to the lack of self-adjoint operators.

\acknowledgements

We wish to thank Abhay Ashtekar and Thomas Thiemann for discussions. 
This work was supported in part by grant NSF-PHY-0968871, funds of the
Hearne Institute for Theoretical Physics, CCT-LSU and Pedeciba. This
publication was made possible through the support of a grant from the
John Templeton Foundation. The opinions expressed in this publication
are those of the author(s) and do not necessarily reflect the views of
the John Templeton Foundation.

\end{document}